\documentstyle[11pt]{article}
\newcommand{\CMP}[1]{{\em Commun. Math. Phys.} {\bf {#1}}}
\newcommand{\JMP}[1]{{\em J.~Math. Phys.} {\bf {#1}}}
\newcommand{\NP}[1]{{\em Nucl.Phys.~B} {\bf {#1}}}

\newcommand{\PL}[1]{{\em Phys. Lett.} {\bf {#1}}}
\newcommand{\PR}[2]{{\em Phys. Rev.} {#1} {\bf {#2}}}

\newcommand{\Ann}[1]{{\em Ann. Phys.} {\bf {#1}}}
\newcommand{\xx}{\!\stackrel {\scriptscriptstyle \times}%
{\scriptscriptstyle \times}\!}
\newcommand{\sst}{\scriptscriptstyle\!}
\newcommand{\half}{\mbox{$\frac{1}{2}$}}

\newcommand{\Lam}{\Lambda}
\newcommand{\dLam}{\Lambda^*}
\newcommand{\Int}{\int_{\Lam}}
\newcommand{\dInt}{\hat{\int}_{\dLam}}

\newcommand{\ddel}{\hat{\delta}}
\newcommand{\Del}{\hat{d}}
\newcommand{\sss}[1]{{\bf {#1}}}

\newcommand{\mmpp}{\stackrel{\scriptscriptstyle -}{\scriptscriptstyle (+)}}

\newcommand{\glt}{\stackrel{\scriptscriptstyle >}{\scriptscriptstyle <}}

\newcommand{\SU}{{\rm SU}}
\newcommand{\U}{{\rm U}}

\newcommand{\eq}{\begin{equation}}
\newcommand{\eqend}{\end{equation}}
\newcommand{\eqa}{\begin{eqnarray}}
\newcommand{\neqa}{\begin{eqnarray*}}
\newcommand{\eqaend}{\end{eqnarray}}
\newcommand{\neqaend}{\end{eqnarray*}}
\newcommand{\nonu}{\nonumber \\ \nopagebreak}
\newcommand{\bma}[1]{\begin{array}{#1}}
\newcommand{\ema}{\end{array}}
\newcommand{\bc}{\begin{center}}
\newcommand{\ec}{\end{center}}

\newcommand{\Ref}[1]{(\ref{#1})}

\newcommand{\dd}{d}

\newcommand{\ee}[1]{\mbox{{\rm e}}^{#1}}
\newcommand{\ii}{{\rm i}}

\renewcommand{\phi}{\varphi}
\newcommand{\sig}{\sigma}
\newcommand{\del}{\delta}
\newcommand{\Om}{\Omega}

\newcommand{\eps}{\varepsilon}

\newcommand{\R}{{\rm I\kern-.2emR}}

\newcommand{\Z}{{\sf Z} \! \! {\sf Z}}

\newcommand{\f}{\frac}

\newcommand{\cD}{{\cal D}}
\newcommand{\cL}{{\cal L}}

\newcommand{\cP}{{\cal P}}
\newcommand{\cH}{{\cal H}}

\newcommand{\ccr}[2]{{[} {#1},{#2} {]} }        
\newcommand{\car}[2]{{\{} {#1},{#2} {\}} }        

\newcommand{\tra}[1]{{\rm tr}({#1})}          
\newcommand{\Tra}[1]{{\rm tr} \left({#1}\right)}          
\newcommand{\normal}[1]{:{#1}:\;}
\newcommand{\Normal}[1]{\vdots{#1}\vdots\;}

\newcounter{saveeqn}
\newcounter{App} 

\newcommand{\alpheqn}{%
\stepcounter{equation}
\setcounter{saveeqn}{\value{equation}}%
\setcounter{equation}{0}%
\renewcommand{\theequation}{\arabic{saveeqn}\alph{equation}} }
\newcommand{\reseteqn}{\setcounter{equation}{\value{saveeqn}}%
\renewcommand{\theequation}{\arabic{equation}} }

\begin{document}
\pagestyle{empty}
\renewcommand{\thefootnote}{\alph{footnote}}
\begin{center}

{\Large \bf QCD$_{\mbox{\small\bf 1+1}}$ with massless quarks and
gauge covariant Sugawara construction}\\
\vspace{1 cm}
{\large Edwin Langmann}\footnote{supported in part by the
"\"Osterreichische Forschungsgemeinschaft" under contract 09/0019.}\\
\vspace{0.2 cm}
{\em Physics Deptartment, Royal Institute of Technology\\100 44 Stockholm,
Sweden}\\
\vspace{.3 cm}
{\large and}\\
\vspace{.2 cm}
{\large Gordon W.  Semenoff}\footnote{supported in part by the
Natural Sciences and Engineering Research Council of Canada.} \\
\vspace{0.2 cm}
{\em Department of Physics, The University of British Columbia\\
Vancouver, B.C., V6T 1Z1, Canada }\\

\end{center}

\setcounter{footnote}{0}
\renewcommand{\thefootnote}{\arabic{footnote}}

\begin{abstract}
We use the Hamiltonian framework to study massless QCD$_{1+1}$, i.e.\
Yang-Mills gauge theories with massless Dirac fermions on a cylinder
(= (1+1) dimensional spacetime $S^1\times \R$) and make explicite the
full, non-perturbative structure of these quantum field theory
models.  We consider $N_F$ fermion flavors and gauge group either
$\U(N_C)$, $\SU(N_C)$ or another Lie subgroup of $\U(N_C)$. In this
approach, anomalies are traced back to kinematical requirements such
as positivity of the Hamiltonian, gauge invariance, and the condition
that all observables are represented by well-defined operators on a
Hilbert space. We also give equal time commutators of the energy
momentum tensor and find a gauge-covariant form of the (affine-)
Sugawara construction.  This allows us to represent massless
QCD$_{1+1}$ as a gauge theory of Kac-Moody currents and prove its
equivalence to a gauged Wess-Zumino-Witten model with a dynamical
Yang-Mills field.

\end{abstract}

\newpage
\pagestyle{plain}
\setcounter{page}{1}

\sss{0. Introduction.}

The path integral formalism provides a very convenient starting point
for perturbative calculations in quantum field theory. Alternatively,
an algebraic framework in the spirit of the Hamiltonian formalism can
be very useful for understanding the non-perturbative structure of
interacting quantum field theories, especially quantum gauge
theories.  In this letter we wish to illustrate this idea in the
simple context of (1+1) dimensions.  We study massless QCD$_{1+1}$,
i.e.\ massless Dirac fermions on spacetime $S^1\times\R$ coupled to a
Yang-Mills (YM) field \cite{HH,M2,LS}, and we demonstrate that in
this case all the mathematical tools and results required to complete
such an algebraic approach do exist (they have been mostly developed
in the context conformal quantum field theory and the representation
theory of the affine Kac-Moody algebras (= current algebras on $S^1$)
and the Virasoro algebra (for references close to the spirit of the
present paper see \cite{GO,CR,GLrev}, for a recent discussion of the
history of the subject we refer to \cite{Halpern}).  Using the
latter, we outline of a simple, non-perturbative and rigorous
construction of these quantum gauge theory models in terms of
operators on a Hilbert space.

We restrict ourselves to the massless case for simplicity, mainly
because representations of the affine Kac-Moody algebras in Fock
spaces of fermions with mass $m>0$ are more complicated and less
understood than for $m=0$ \cite{CR}.

In general, we allow for $N_F$ fermion flavors and a gauge group
$H=\U(N_C)$, $\SU(N_C)$, or another Lie subgroup of $\U(N_C)$.  Put
differently, we use fermions transforming in the fundamental
representation of the group $G=\U(N_C)\times U(N_F)$ and gauge the Lie
subgroup $H$ of $G$.  To simplify our notation, we first concentrate
on the special case $N_F=1$ and $H=G=U(N_C)$, and then discuss the
modifications required for the general case in Paragraph 7.

We note that a rigorous construction of 2 dimensional QCD has also
been given by Klimek and Kondracki \cite{KK}. In contrast to our
approach, they study the model in 2 Euclidean dimensions and use
methods from constructive field theory (see e.g.\ \cite{Riv}) which are
close in spirit to the path integral formalism. We believe that our
results here provide an example that Hamiltonian methods can provide a
powerful alternative to these methods (of course, at the moment the
latter are much further developed than the former \cite{Riv}).

\sss{1. Preliminaries.} To fix our notation and summarize the
algebraic structure of the model, we first recall the canonical
formalism for massless QCD$_{1+1}$ on the semiclassical level (= on
the unphysical Hilbert space, no filled Dirac sea).

Let $G=\U(N_C)$ be the structure group of the YM-field and $T^a$ the
generators of the Lie algebra $g$ of $G$ in the fundamental
representation of $\U(N_C)$ obeying\footnote{$\ccr{\cdot}{\cdot}$ is the
commutator, $*$ and $\tra{\cdot}$ the adjoint and the trace of
$N\times N$-matrices, respectively; repeated indices are summed over
throughout unless stated otherwise}
\newcommand{\tu}{\tau}
\eq
\label{b.1}
\ccr{T^a}{T^b}=\ii\lambda^{ab}_{\;\;\; c}T^c,\quad (T^a)^*=T^a,\quad
\tra{T^aT^b}=\tu^{ab}
\eqend
with $(\tu^{ab})$ an invertible matrix (`metric tensor in color
space'). Denoting as $(\tu_{ab})$ the inverse matrix of $(\tu^{ab})$,
we also introduce $T_a\equiv
\tu_{ab}T^b$, and for $X\in g$ we write $X=X_a T^a= X^aT_a$ where
$X_a=\tra{XT_a}=\tu_{ab}X^b$ etc. Then $\tra{XY}=X_aY^a$ for $X,Y\in
g$.

With $A_\nu\equiv A_\nu^a T_a$ the YM-field and $\psi,\bar{\psi}\equiv
\psi^*\gamma^0$ the fermion fields, the Lagrangian density for
massless QCD$_{1+1}$ is
\eq
\label{1}
\cL_{\rm QCD} = -\f{1}{4 e^2}\Tra{F_{\mu\nu}F^{\mu\nu}}
-\ii \bar{\psi}\gamma^\nu D_\nu\psi
\eqend
with $D_\nu\psi=(\partial_\nu + \ii A_\nu)\psi$, $e$ the coupling
constant, $\partial_\nu\equiv\partial/\partial x^\nu$, $x^0 =t\in\R$
time, $x^1=x\in \Lambda\equiv [-L/2,L/2]$ the spatial coordinate, and
$
\label{2}
F_{\mu\nu}=\partial_\mu A_\nu - \partial_\nu A_\mu + \ii
\ccr{A_\mu}{A_\nu}
$ the YM field strength\footnote{$\mu,\nu\in\{0,1\}$ are space-time
indices; our metric is $g_{\mu\nu}=diag(1,-1)$, and the anti-symmetric
tensor $\eps^{\mu\nu}$ with $\eps^{01}=1$}.
More explicitly, $\gamma^\nu\equiv\gamma^\nu_{\sigma\sigma'}$,
$T^a\equiv T^a_{AB}$, and $\psi^{(*)}\equiv
\psi^{(*)}_{\sig,A}$, $\sig,\sig'\in\{1,2\}$ and
$A,B\in\{1,2,\ldots,N\}$ are spin and color indices, respectively.  To
be specific, we choose the Dirac matrices as $\gamma^0=\sigma_1$,
$\gamma^1=\ii\sigma_2$, $\gamma_5\equiv -\gamma^0\gamma^1=\sigma_3$
($\sigma_i$ the Pauli matrices as usual).

With the usual canonical procedure \cite{Su} we get the momenta
$\Pi_a^\nu$ conjugate to $A^a_\nu$, {\em viz.}\footnote{the symbol
`$\simeq$+ means weak equality (constraint)} $\Pi_a^0(x)\simeq 0$ and
$\Pi_a^1(x)=\f{1}{e^2}(F_{01})_a(x)$, and the following canonical
(anti-) commutator relations (C(A)CR)\footnote{$\car{\cdot}{\cdot}$ is the
anti-commutator}
\eqa
\label{cacr}
\ccr{\Pi^\mu_a(x)}{A_{\nu}^b(y)} &=&
\ii\delta^\mu_\nu\delta_a^b \del(x-y)\nonu
\car{\psi_{\sig,A}(x)}{\psi^*_{\sig',B}(y)} &=&
\del_{\sig\sig'}\del_{AB}\del(x-y)
\eqaend
etc. as usual.  Moreover, the resulting Hamiltonian is\footnote{we
assume periodic boundary conditions for the YM fields}
\eq
\label{4}
H_{\rm QCD} = H^{(0)}_{\rm F} +\Int\dd{x}\,\Tra{\f{e^2}{2}\Pi_1(x)^2 +
A_1(x)j(x)- A_0(x)G(x)}
\eqend
where we introduced the free fermion Hamiltonian
\eq
\label{5}
H^{(0)}_{\rm F} \equiv
\Int\dd{x}\,\psi^*(x)\gamma_5(-\ii\partial_1)\psi(x) ,
\eqend
the fermion currents
\eqa
\label{6}
\rho^a(x) &\equiv& \psi^*(x) T^a \psi(x) \nonu
j^a(x) &\equiv& \psi^*(x) \gamma_5 T^a \psi(x)
\eqaend
and $j=j_aT^a$  etc., and the Gauss law
operators
\eq
\label{7}
G(x) \equiv -D_1\Pi_1(x) + \rho(x)
\eqend
where
$
(D_\nu X)_a\equiv \partial_\nu X_a + \ii\ccr{A_\nu}{X}_a=
\partial_\nu X_a - \lambda_{abc} A_\nu^b X^c .
$

The primary constraint $\Pi_a^0\simeq 0$ implies the secondary constraint
$\ccr{\Pi_a^0(x)}{H_{\rm QCD}}=-\ii G_a(x) \simeq 0$ (Gauss' law).
One also gets $\ccr{G_a(x)}{H_{\rm QCD}} =
-\ii\lambda_{abc}A^b_0(x)G^c(x)\simeq 0, $ hence there are no
tertiary constraints.

{}From \Ref{1} we deduce that the vector-- and the axial fermion
currents,
$J_a^\nu=\bar\psi\gamma^\nu T_a\psi$ and
$(J_5^\nu)_a=\bar\psi\gamma^\nu
\gamma_5 T_a\psi$,
obey on the semi-classical level the equations of motion $D_\nu
J^\nu=D_\nu J_5^\nu =0$.

We also obtain the symmetric and gauge invariant energy-momentum tensor
\eq
\Theta^{\mu\nu}=\f{1}{e^2}\Tra{\mbox{$\f{1}{4}$}
g^{\mu\nu}F^{\alpha\beta} F_{\alpha\beta} -
F^{\mu\alpha}F^\nu_{\;\;\alpha}}-
\mbox{$\f{\ii}{2}$}\bar\psi\left[\gamma^\mu D^\nu + \gamma^\nu D^\mu
\right]\psi
\eqend
which is derived on-shell, i.e.\ by taking into account the eqs.\ of
motion $D_\mu F^{\mu\nu} + e^2 J^\nu=0$ and $\gamma^\mu D_\mu\psi=0$.
We also use light-cone coordinates, i.e.\
\eq
\label{rhopm}
J^\pm(x)\equiv \half\left(\rho(x) \mp j(x)\right)
\eqend
and similarly for $\Theta$. We write
\alpheqn
\eqa
L^\pm &\equiv& \mp \Theta^{\pm\pm} =
-\ii\psi^*\half(1\mp\gamma_5)D_1\psi \equiv
 L^\pm_0 + \Tra{A_1 J^\pm} \\
M&\equiv& \f{1}{e^2}\Theta^{+-} = \mbox{$\f{1}{4}$}\Tra{\Pi_1^2}
=\f{1}{e^2}\Theta^{-+},
\eqaend
\reseteqn
where again we used the equation of motion for $\psi$.

\sss{2. Fourier Transformation.}
In the following we find it convenient to work in Fourier space.
Having in mind the thermodynamic limit $L\to\infty$ at the end of our
construction, we use the following suggestive notation: Fourier space
is
$\dLam\equiv \left\{\left. k=\f{2\pi}{L} n \right| n\in\Z\right\}$,
and for functions $\hat f$ on $\dLam$ we write
$\dInt \Del k \hat f(k) \equiv \sum_{k\in\dLam} \f{2\pi}{L} \hat
f(k)$ so that $\del(x-y) = \dInt \f{\Del k}{2\pi} \ee{\ii k(x-y)}$.
Then the appropriate $\del$-function on $\dLam$ is $\ddel(k-q) \equiv
\del_{k,q}L/2\pi =\Int \f{\dd x}{2\pi} \ee{ -\ii x(k-q)}$ for
$k,q\in\dLam$.

For the Fourier transformed operators we use the following conventions
($k\in\dLam$),
\alpheqn
\eq
\label{10a}
\hat \psi^{(*)}(k) =
\Int \f{\dd{x}}{\sqrt{2\pi}} \psi^{(*)}(x)
\ee{\mmpp \ii kx}
\eqend
(to simplify our notation, we use periodic boundary conditions for the
fermions; it is trivial to modify our eqs.\ so as to allow for
anti-periodic boundary conditions),
\eq
\label{A1}
\hat A_\nu(k) = \Int \f{\dd{x}}{2\pi}
A_\nu(x) \ee{-\ii kx},
\eqend
and in all other cases
\eq
\label{other}
\hat X(k) = \Int \dd{x} X(x) \ee{-\ii kx}\quad \mbox{ for
$X=\Pi_\nu,G,\rho,j,J^\nu,J_5^\nu$, and $\Theta$.}
\eqend
\reseteqn
For convenience of the reader, we write down the non-trivial C(A)CR in
Fourier space,
\eqa
\label{fcacr}
\ccr{\hat \Pi_\mu^{a}(k)}{\hat A_\nu^{b}(q)} &=&
\ii g_{\mu\nu}\tu^{ab}\ddel(k+q) \nonu
\car{\hat \psi_{\sigma,A}(k)}{\hat \psi^*_{\sigma',B}(q)}
&=& \delta_{\sigma\sigma'}\delta_{AB}\ddel(k-q) \quad \forall
k,q\in\dLam .
\eqaend

\sss{3. Filling the Dirac Sea.} Our approach is in the
spirit of the algebraic approach to quantum field theory
\cite{HK} where the non-trivial aspects of quantum field theory (as
compared to quantum mechanics) arise due to the existence of unitarily
inequivalent representations of quantum field algebras \cite{BR2}.
The essential {\em physical requirement} selecting the appropriate
representation is {\em positivity of the Hamiltonian on the physical
states}.  The crucial simplification in (1+1) (and not possible in
higher) dimensions is that one can use a quasi-free representation
\cite{CR} for the fermion field operators corresponding to ``filling
up the Dirac sea'' associated with the {\em free} fermion Hamiltonian
$H^{(0)}_{\rm F}$, and for the YM operators one can use the naive
Schr\"odinger representation.  At this point, we have to take this as
an assumption to be checked at the end of the construction. However,
this assumption is plausible due to the facts that, (i) quasi-free
representations for fermion fields in different {\em external}
YM field are unitarily equivalent in (1+1) dimensions
\cite{M1}, (ii) the YM field on a cylinder has only a finite
number of {\em physical} degrees of freedom (namely the eigenvalues of the
parallel transporter $\cP\exp{(\ii\Int \dd{x} A_1(x))}$, see e.g.\
\cite{LS}), and as all representations of a finite number of quantum
degrees of freedom are unitarily equivalent (von Neumann's theorem), the
simplest representation for the YM field algebra should do.

We therefore construct a representation of the C(A)CR algebra given
above on a Hilbert space $\cH$ which is a tensor product of a YM and a
fermion Hilbert space, $\cH=\cH_{\rm YM}\otimes\cH_{\rm F}$, with
$\cH_{\rm YM}$ the usual Hilbert space of functionals of $\hat
A_\nu^a(k)$ with $\hat\Pi^\nu_a(k)=\f{L}{2\pi}\ii\partial/\partial
\hat A_\nu^a(-k)$, and $\cH_{\rm F}$ the Fermion Fock space with vacuum
$\Omega_{\rm F}$ such that
\eqa
\label{11a}
 \half(1+\gamma_5) \hat\psi(k)\Omega_{\rm F} =
\half(1-\gamma_5)\hat\psi^*(k)\Omega_{\rm F} =0\quad \forall k>0
\nonu
 \half(1+\gamma_5)\hat\psi^*(k)\Omega_{\rm F} =
\half(1-\gamma_5)\hat\psi(k)\Omega_{\rm F} = 0 \quad
\forall k\leq 0
\eqaend
(by abuse of notation, we do not distinguish the quantities introduced
on the semi-classical level in the last Paragraph from the
well-defined operators representing them on $\cH$).  It is well-known
that the presence of the Dirac sea requires normal-ordering
$\normal{\cdots}$ of the fermion bilinears, hence $\tilde H^{(0)}_{\rm
F}=\dInt\Del q\normal{q\hat\psi^*(q)\gamma_5\hat\psi(q)}$ (which is
positive by construction \cite{GLrev}), and similarly for $\tilde
J_a^\pm(k)$ and $\tilde L_0^\pm(k)$, where the tilde indicates {\em
normal ordering with respect to the free fermion vacuum $\Omega_{\rm F}$}.
This modifies their naive commutator relations following from the CAR
\Ref{fcacr} as Schwinger terms show up \cite{GO,GLrev} (for a
mathematical rigorous discussion of this construction of fermion
bilinears in the presence of a Dirac sea and how normal ordering leads
to Schwinger terms, see \cite{CR}).  In our case
\cite{GO,GLrev},
\alpheqn
\eq
\label{12}
\ccr{\tilde J_a^\pm(k)}{\tilde J_b^\pm(q)}=
\ii\lambda_{ab}^{\;\;\; c}\tilde J_c^\pm(k+q) \mp
k\ddel(k+q) \tu_{ab}
\eqend
and
\eq
\label{12a}
\ccr{\tilde L_0^\pm(k)}{\tilde L_0^\pm(q)} = (k-q)\tilde L_0^\pm(k+q) \mp
 \mbox{$\f{N_C}{6}k\left(k^2-\left(\f{2\pi}{L}\right)^{\sst 2}\right)
\ddel(k+q)$}
\eqend
with the second terms on the r.h.s. of \Ref{12} and \Ref{12a} the
Kac-Moody and Virasoro cocycles, respectively \cite{GO}.  Moreover,
\eq
\label{12b}
\ccr{\tilde L_0^\pm(k)}{\tilde J^\pm_a(q)} = -q\tilde J^\pm_a(k+q)
\eqend
\reseteqn
with no Schwinger term arising here. Note that these relations are
exactly the ones of the semi-direct product of an affine Kac-Moody
algebra and the Virasoro algebra playing a prominent role in conformal
field theory.

We can now write the Gauss' law as
$\hat G_a(k) = -\widehat{(D_1 \Pi_1)}_a(k) + \tilde\rho_a(k) \simeq 0$
where $ \widehat{(D_1 \Pi_1)}_a(k) = \ii k(\hat\Pi_1)_a(k) -
\lambda_{abc}\dInt\Del q \hat A_1^b(k+q)\hat\Pi_1^c(-q) , $ so eqs.
\Ref{12} imply that
\[
\ccr{\hat G_a(k)}{\tilde J_b^\pm(q)} =
\ii\lambda_{ab}^{\;\;\;c}\tilde J_c^\pm(k+q)
\mp k\ddel(k+q)\tu_{ab}.
\]
Thus the presence of the Schwinger terms implies that these fermion
currents no longer have the classical commutator relations with the
Gauss' law generators and therefore do not transform covariantly under
gauge transformations.

To restore gauge covariance and obtain fermion currents having
canonical transformation properties (without Schwinger terms), we note
that $
\ccr{\hat G_a(k)}{(\hat A_1)_b(q)} = - k\ddel(k+q)\tu_{ab} +
\ii\lambda_{abc}\hat A_1^c(k+q),
$
hence the operators
\eq
\label{16}
\hat J^\pm (k) \equiv \tilde J^\pm (k)
 \mp \hat{A_1}(k)
\eqend
obey the desired relations
\eq
\ccr{\hat G_a(k)}{\hat J_b^\pm (q)} =
\ii\lambda_{ab}^{\;\;\;c}\hat J_c^\pm (k+q).
\eqend
Similarly, the naive energy-momentum components
$
\tilde L^\pm(k) = \tilde L^\pm_0(k) +
\dInt \Del q \, \Tra{\hat A_1(k+q)\tilde J^\pm(-q)}
$
are not gauge invariant, $\ccr{\hat G^a(k)}{\tilde L^\pm_0(q)} =
 \mp k\hat A_1^a(k+q)$, but obviously there are unique polynomials in
$\hat A_1$ which can be added to make them gauge invariant,
\eq
\label{Lpm}
\hat L^\pm(k) = \tilde L^\pm_0(k) +
\dInt \Del q \,
\Tra{\hat A_1(k+q)\tilde J^\pm(-q) \mp \half \hat A_1(k+q)\hat A_1(-q)}.
\eqend
\reseteqn
Recalling that normal ordering is only unique up to finite terms, it is
natural to regard the $\hat J_a^\pm (k)$ and $\hat L^\pm(k)$ as the
currents and energy-momentum components obtained by a {\em gauge
covariant normal ordering} preserving the transformation properties
under gauge transformations.

To construct the full energy momentum tensor --- especially the
Hamiltonian --- we also need the operators
\eq
\label{M}
\hat M(k)= \dInt \Del q \, \f{1}{8\pi}\Tra{\hat \Pi_1(k+q)\hat \Pi_1(-q)} .
\eqend
At this point a technical difficulty arises: the operators $\hat
L^\pm(k)$ and $\hat M(k)$ do not have a common, dense invariant domain
of definition in $\cH$ (we recall that the sum of two unbounded
Hilbert space operators can be defined directly only if these
operators have such a domain). It is, however, possible to define the
operators $\Normal{\hat L^\pm(k)\pm e^2 \hat M(k)}$ --- and therefore
the components $\hat\Theta^{00}(k)$ and $\hat\Theta^{01}(k)$ of the
energy momentum tensor --- by normal ordering $\Normal{\cdots}$ the
YM field operators with respect to the YM vacuum $\Om_{\rm YM}$
obeying
\eq
\left( \f{e}{\sqrt{4\pi}}\ii\hat\Pi_1(k) + \hat
A_1(k)\right)
\Om_{\rm YM}=0\quad\forall k\in\dLam.
\eqend
Note that $\Om_{\rm YM}$ is just the ground state of the free YM
Hamiltonian
\eq
H^{(0)}_{\rm YM}=\Normal{\dInt\Del k \, \Tra{ \f{e^2}{4\pi}
\hat \Pi_1(k)\hat \Pi_1(-k) + \hat A_1 (k)\hat A_1(-k)}}.
\eqend

Especially, we get the gauge invariant Hamiltonian $H_{\rm QCD}=
\hat\Theta^{00}(0) -G(A_0)$, or equivalently
\eq
\label{17a}
H_{\rm QCD} = \tilde H^{(0)}_{\rm F} + H^{(0)}_{\rm YM}+ \dInt\Del k
\,
\Tra{\hat A_1(k)
\tilde j(-k) - \hat A_0(k)\hat G(-k) }.
\eqend
We see that, similarly as in the Schwinger model \cite{Manton}, there
is an additional `gluon mass term' resulting from gauge invariant
normal ordering.

\sss{4. Observable Algebra Relations.}
For the covariant currents we get the following commutators,
\eq
\label{JJ}
\ccr{\hat J_a^\pm(k)}{\hat J_b^\pm(q)}=
\ii\lambda_{ab}^{\;\;\; c}\hat J_c^\pm(k+q) \mp
\hat S_{ab}(k,q)
\eqend
with the Schwinger term
\eq
\hat S_{ab}(k,q) = k\ddel(k+q) \tu_{ab} -\ii\lambda_{abc}\hat A_1^c(k+q).
\eqend
It is natural to regard the latter as the {\em gauge covariant form of
the Kac-Moody cocycle}, and we can represent it in explicitly
covariant form as $\hat S_{ab}(k,q) =
-\ccr{\hat\Pi_a^0(k)}{\widehat{(D_1A_0)}_b(q)}$.

Moreover, we get
\eq
\label{LpmLpm}
\ccr{\hat L^\pm(k)}{\hat L^\pm(q)} = (k-q)\hat L^\pm(k+q)\mp
 \mbox{$\f{N_C}{6}k\left(k^2-\left(\f{2\pi}{L}\right)^{\sst 2}\right)
\ddel(k+q)$}
\eqend
and
\eq
 \ccr{\hat L^\pm(k)}{\hat J^\pm_a(q)} = -q\hat J^\pm_a(k+q) -
\widehat{\ccr{A_1}{J^\pm}}_a(k+q).
\eqend

{\em Remark:} For some readers it might appear that there is a
contradiction between the existence of Virasoro and current algebras and the
model not being conformally invariant.  However, as the Hamiltonian of the
model is {\em not} equal to $-\hat L^+(0)+\hat L^-(0)$, these algebras do
not correspond to symmetries of the model which could be used to answer
dynamical questions, but they only provide interesting (non-conserved!)
observables.  It might be useful to recall an analogous situation in
quantum mechanics: the harmonic oscillator operators $a$ and $a^*$ obeying
$[a,a^*]=1$ exist for every QM model, but only for the harmonic oscillator with
the
Hamiltonian $\propto a^*a$ their algebraic properties become powerful for
answering dynamical questions.

\sss{5. Equations of Motion.} Using \Ref{17a} and the algebraic
relations above it is straightforward to work out the equations of
motion for all observables of the model. For example, we obtain
$\ccr{H_{\rm QCD}}{\hat J^\pm (k)} =
\mp\ii\widehat{(D_1 J^\pm)}(k) + \widehat{\ccr{A_0}{J^\pm}}(k)
\pm \ii \f{e^2}{2\pi}\hat\Pi(k)$.
Using $\partial_0 J^\pm + \ii\ccr{H_{\rm QCD}}{J^\pm}=0$, and
transforming to position space, we can write this as $D_0 J^\pm \pm
D_1 J^\pm = \pm
\f{e^2}{2\pi}\Pi_1$.  Noting that
$\Pi_1=-\Pi^1=-\f{1}{2e^2}\eps^{\mu\nu}F_{\mu\nu}$ and $J^0=J^+
+J^-=-J^1_5$, $J^1=-J^+ +J^- = -J^0_5$, this can be written as
\alpheqn
\eqa
D_\nu J^\nu &=& 0\\ \label{anomaly} D_\nu J_5^\nu &=&
\f{1}{2\pi}\eps^{\mu\nu}F_{\mu\nu} .
\eqaend
\reseteqn
The second of these eqs.\ shows a covariant axial anomaly.

Evaluating $\partial_0 \Pi_1(k) +
\ii\ccr{H_{\rm QCD}}{\Pi_1}=0$ we obtain $D_0 \Pi_1 - j=0$. Together
with the Gauss' law --- which we can write as $D_1\Pi_1 - \rho\simeq
0$ --- this comprises the usual equations of motion of the YM field
\eq
D_\mu F^{\mu\nu} + e^2 J^\nu\simeq 0
\eqend
identical to those obtained on the semi-classical level. Using this and
eq.\ \Ref{anomaly} rewritten as $D_0 j - D_1\rho = -
\f{e^2}{\pi}\Pi_1$, we obtain
$D_0^2\Pi_1 - D_1^2\Pi_1 \simeq - \f{e^2}{\pi}\Pi_1$, or equivalently
\eq
\label{KG}
D_\nu D^\nu \Pi_1 + \f{e^2}{\pi}\Pi_1\simeq 0
\eqend
generalizing the Klein-Gordon equation one has in the Abelian case
\cite{Manton}. Equations \Ref{anomaly} and \Ref{KG} have also been
obtained by by Sorensen and Thomas \cite{Sorensen} in a path integral
approach.

\sss{6. Bosonization.}
The celebrated (affine-) Sugawara construction  allows to
write the free Virasoro generators $\tilde L_0^\pm$ in terms of the
Kac-Moody currents $\tilde J^\pm$,
\eq
\label{sugawara}
\tilde L^\pm_0(k) = \mp\half\dInt \Del q\, \xx
\Tra{\tilde J^\pm(k+q) \tilde J^\pm(-q)}\xx
\eqend
(see e.g. \cite{GO}; for $k=0,\pm 2\pi/L$ this was already given in
\cite{H1}) with normal ordering $\,\xx\!\tilde J_a^\pm(k) \tilde
J_b^\pm(q)\!\xx
\;\; \equiv\tilde J_b^\pm(q) \tilde J_a^\pm(k)$ for $k\glt q$ and
$\tilde J_a^\pm(k)\tilde J_b^\pm(q)$ otherwise \cite{GO} (note that
$\tilde J^+(-k)\Omega_{\rm F}=\tilde J^-(k)\Omega_{\rm F}=0$ $\forall k>0$).
Combining this with eqs.\ \Ref{Lpm} and \Ref{16}, we observe that the terms
involving $\hat A_1$ can be arranged such that
\eq
\hat L^\pm(k) = \mp\half\dInt \Del q \, \xx
\Tra{\hat J^\pm(k+q) \hat J^\pm(-q)}\xx
\eqend
Thus the Virasoro generators $\hat L^\pm(k)$ are obtained simply by
replacing the non-covariant currents on the r.h.s. of eq.\
\Ref{sugawara} by the covariant ones!  It is natural to regard this as
the {\em gauge covariant version of the Sugawara construction}.

Especially, we get the Hamiltonian of the model in the following form
\eq
H_{\rm QCD} = \half\Normal{\dInt \Del k \, \Tra{\xx
\left( \hat J^+(k)\hat J^+(-k)
 + \hat J^-(k)\hat J^-(-k) \right)\xx +
\f{e^2}{2\pi}\hat\Pi_1(k)\hat\Pi_1(-k)-\hat A_0(k) \hat
G(-k) }}.
\eqend
It is now manifestly positive definite on the physical Hilbert space
where $\hat G(k)\simeq 0$ (note that $\hat J^\pm(-k)=\hat J^\pm(k)^*$)
thus justifying our choice of representation of the field algebra.

It is easy to see that this Hamiltonian is identical with the one
obtained from a gauged Wess-Zumino-Witten model with dynamical gauge
field\footnote{i.e.\ the Lagrangian has a term
$-\f{1}{4g^2}\Tra{F_{\mu\nu}F^{\mu\nu}}$ in addition to what is
usually referred to as gauged WZW model \cite{Bowcock}} (see e.g.\
\cite{Bowcock}) and a gauge group $\U(N_C)$ equal the flavor group and
coupling constant $g=e/\sqrt{\pi}$ (to make this explicit, one has to
rescale $\sqrt\pi A_\nu(x)\to A_\nu(x)$). This equivalence has been
known from the path integral approach \cite{Affleck,Sonnenschein} (see
also
\cite{GWZW}).

\sss{7. General Case.} Our present approach can be immediately
generalized to the case with a gauge group $H$ being a Lie subgroup of
$\U(N_C)$ and $N_F$ fermion flavors. In this case, we have fermions
transforming under the fundamental representation of
$G=\U(N_C)\times\U(N_F)$ with the gauge group $H$ a subgroup of $G$.
Labeling the generators of the Lie algebra of $G$ such that $T_a$ for
$1\leq a\leq dim(H)$ span the Lie algebra of $H$ and $\tra{T^a T^b}=0$
for $a\leq dim(H), b>dim(H)$, we have $A_1=\sum_{a=1}^{dim(H)} A_1^a
T_a$ and similarly for $F_{\mu\nu}$, $G$. With that all equations and
the discussion from Paragraphs 1--4 essentially remain the same.

For the bosonization, it is most convenient to use the $G/H$ coset
construction \cite{GO},
\eq
 \tilde L_0^\pm = (\tilde L_0^\pm)_G = (\tilde L_0^\pm)_{H} + (\tilde
L_0^\pm)_{G/H}
\eqend
which implies a similar equation for $\hat L^\pm = \hat L^\pm_G$.
Then obviously $\hat L^\pm_{G/H}=(\tilde L_0^\pm)_{G/H}$, i.e.\ the
coset Virasoro generators are completely decoupled from the gauge
field, and the discussion of gauge covariant normal ordering etc.
above applies to $\hat L^\pm_H$ only.

Especially the Hamiltonian of the model is $H_{\rm QCD}=H_H +
H_{G/H}$ where $H_{G/H}=(\tilde H^{(0)}_{\rm F})_{G/H}$ is completely
decoupled from all YM fields, commutes with $H_H$, and is identical
to the one one gets in the free fermion case \cite{GO}.  The
non-trivial interactions of the fermions with the YM field are
completely contained in $H_H$.  Thus we can write
$\cH=\cH_H\otimes\cH_{G/H}$ with all non-trivial dynamics occurring
on $\cH_H$, and $\cH_{G/H}$ provides the superselection sectors of
the model (cf.\ also \cite{Affleck}).

\sss{8. Technicalities.} To complete the construction of massless QCD$_{1+1}$
and make it mathematically rigorous, one has to establish several
technical properties. Firstly (as most of the operators of the model
are unbounded), one has to prove that there is a common, dense,
invariant domain $\cD\subset\cH$ for all operators of interest so that
the commutator relations given above are well-defined on $\cD$.  In
fact, this can be proven for the Hamiltonian $H_{\rm QCD}$ and all other
operators considered above except the energy momentum components
$\hat\Theta^{11}(k)$ (see also the discussion in Paragraph 3; taking
also the latter into account makes things slightly more complicated
\cite{CL}).

Secondly, one has to prove that all observables of the model,
especially the Hamiltonian and the (smeared) Gauss law generators, are
represented not only by symmetric but in fact self-adjoint operators
on $\cH$ \cite{RS1,RS2}.  Finally, one would like to establish that
the thermodynamic limit $L\to\infty$ is well-defined and leads to a
relativisticly invariant theory. The proof of these results can be
done by using techniques developed in \cite{CR,GL2} in combination
with results summarized in \cite{Simon} and will appear elsewhere
\cite{CL}.

\sss{9. Final Comments.}
Recently an interesting reformulation of QCD on spacetime $\R\times\R$
in terms of a gauge invariant, bilocal master field was given and used
as a starting point for a systematic semi-classical approximation
\cite{Rajeev} (see also \cite{DMW}).  It would be interesting
understand this reformulation in our framework (technically this is
more complicated due to the presence of the physical YM degrees of
freedom on spacetime $S^1\times\R$). Alternatively one can
eliminate the gauge degrees of freedom by `solving the Gauss' law'
\cite{LS2}. For massless QCD$_{1+1}$ this results in a theory of
interacting Kac-Moody currents $J^\pm$ coupled to a finite number of
quantum mechanical degrees of freedom, the latter
representing the physical YM degrees of freedom \cite{LS}.
{}From a mathematical point of view this gauge fixing procedure is
quite delicate, and it would be important to get a deeper
understanding, e.g. in the general framework of \cite{HG}.  Work in
this direction is in progress
\cite{CL}.

There are several deep reasons preventing a straightforward extension
of our construction to higher dimensions.  Most importantly, a YM
field there has also an infinite number of physical degrees of
freedom, and choosing the appropriate representation of the YM field
algebra is a highly non-trivial problem, even for pure YM theory.
Moreover, in higher dimensions the physical representations for
fermions interacting with different external, static YM-fields are not
unitarily equivalent and gauge transformations cannot be implemented
by unitary operators in the fermion sector but only by sesquilinear
forms \cite{MR,L}.  This suggests that the observable algebra of QCD in
higher dimensions at fixed, sharp time does not allow for a reasonable
Hilbert space representation, hence a standard Hamiltonian formalism
might be too narrow a framework for higher dimensional quantum gauge
theories.  There is, however, a natural generalization of the theory
of the affine Kac-Moody algebras to (3+1) dimensions \cite{MR,L} which
can be expected to provide a first step to a non-perturbative
understanding of the fermion sector of QCD$_{3+1}$.

\bc
\subsection*{Acknowledgement}
\ec

One of us (E.L.)  would like to thank I. Affleck, A. Carey, A. Hurst,
F. Lenz, N. Manton, S. Rajeev, M.  Salmhofer, E. Seiler, and S. Weber
for useful discussions. He would also like to thank A.  Carey for an
invitation to the University of Adelaide where part of this work was
done and the Erwin Schr\"odinger International Institute in Vienna
for hospitality where this work was finished.



\begin{thebibliography}{99}




\bibitem{HH} Hetrick J. E. and Hosotani Y.,
\PL{230B}, 88 (1989).

\bibitem{M2} Mickelsson J., \PL{242B}, 217 (1990).

\bibitem{LS} Langmann E. and Semenoff G. W., {\em Phys. Lett.  B}
{\bf 303}, 303 (1993).

\bibitem{GO} Goddard P. and Olive D.,
{\em Int. J. Mod. Phys} {\bf A1}, 303 (1986).

\bibitem{CR} Carey A.~L. and Ruijsenaars S.~N.~M.,
{\em Acta Appl. Mat.} {\bf 10}, 1 (1987).

\bibitem{GLrev} Grosse H. and Langmann E.,
{\em Int. Jour. of Mod. Phys.} {\bf 21}, 5045 (1992).

\bibitem{Halpern} Halpern M.B., {\em Recent progress in irrational
conformal field theory}, proceedings of ``String 1993'', Berkley,
{\tt hep-th/9309087} (to be published in World Scientific).

\bibitem{KK} Klimek S. and Kondracki, \CMP{113}, 389 (1987).

\bibitem{Riv} Rivasseau V.: {\em From Perturbative to Constructive
Renormalization}, Princeton University Press, Princeton, N.Y. (1991).

\bibitem{Su} Sundermeyer K., {\em Constrained Dynamics},
Lecture Notes in Physics 169, Springer Verlag (1982).

\bibitem{HK} Haag R. and Kastler D., \JMP{7}, 848 (1964).

\bibitem{BR2} Bratteli O. and Robinson D.~W.,
{\em Operator Algebras and Quantum Statistical Mechanics. Vol.2}.
Berlin, Springer 1981.

\bibitem{M1} Mickelsson J.: {\em Current Algebras and Groups},
Plenum Monographs in Nonlinear Physics, Plenum Press (1989).

\bibitem{Manton} Manton N. S., \Ann{159}, 220 (1985).

\bibitem{Sorensen} Sorensen C. and Thomas G. H.,
{\em Phys. Rev.} {\bf D 21}, 1625  (1980).

\bibitem{H1} Bardakci K. and Halpern M. B., \PR{D 3}, 2493
(1971).

\bibitem{Bowcock} Bowcock P., \NP{316}, 80 (1989).

\bibitem{Affleck} Affleck I., \NP{265} [FS15], 448 (1986).

\bibitem{Sonnenschein} Frishman Y. and Sonnenschein J., {\em Phys.
Rep.} {\bf 223}, 309 (1993).

\bibitem{GWZW} Brown L. S. and Nepomechie, {\em Phys. Rev.} {\bf D
35}, 3239 (1987)

\bibitem{RS1} Reed R., and Simon B.:
{\em Methods of Modern Mathematical Physics I. Functional Analysis},
Academic Press, New York (1968)

\bibitem{RS2} Reed R., and Simon B.: {\em Methods of Modern
Mathematical Physics II. Fourier Analysis, Self-Adjointness}, Academic
Press, New York (1975).

\bibitem{GL2} Grosse H. and Langmann E., {\em Jour. Math.
Phys.} {\bf 33}, 1032 (1992).

\bibitem{CL} Carey A. L., Hurst C. A.,  and Langmann E., in preparation.

\bibitem{Simon} Simon B.: {\em Quantum Mechanics for Hamiltonians
Defined as Quadratic Forms}, Princeton University Press, Princeton,
New Jersey (1971).

\bibitem{Rajeev} Rajeev S. G., Quantum Hadrodynamics in two
dimensions, {\tt hep-th/9401115} (to appear in {\em Int. Jour. of Mod.
Phys.}).

\bibitem{DMW} Dhar A., Mandal G., and Wadia S. R., {\em String field
theory of two dimensional QCD: a realization of $W_\infty$ algebra},
{\tt hep-th/9403050}.

\bibitem{LS2} Langmann E. and Semenoff G. W., {\em Phys. Lett. B} {\bf
296}, 117 (1992).

\bibitem{HG} Grundling H. and Hurst C. A., \CMP{119}, 75 (1988).

\bibitem{MR} Mickelsson J. and Rajeev S. G., \CMP{116}, 400 (1988).

\bibitem{L} Langmann E., \CMP{162}, 1 (1994).

\end{thebibliography}
\end{document}